%% file: main.tex
\documentclass[journal=jctcce,manuscript=article]{achemso}
\usepackage[T1]{fontenc}

\usepackage{graphicx}
\usepackage{booktabs}
\usepackage{array}
\usepackage{amsmath}
\usepackage{url}

\newcommand{\Eh}{\ensuremath{E_\mathrm{h}}}
\newcommand{\quickflag}{\texttt{-{}-quick}}
\newcolumntype{L}[1]{>{\raggedright\arraybackslash}p{#1}}
\title{FastCREST: GPU-Accelerated CREST Conformer Searches with Consistent Energies and Gradients}

\author{Supantha Dey}
\altaffiliation{These authors contributed equally to this work.}
\affiliation{Department of Chemical and Biological Engineering, Iowa State University, Ames, Iowa 50011, United States}
\alsoaffiliation{Interdepartmental Genetics and Genomics Program, Iowa State University, Ames, Iowa 50011, United States}
\author{Nilusha L Kariyawasam}
\altaffiliation{These authors contributed equally to this work.}
\affiliation{Department of Chemical and Biological Engineering, Iowa State University, Ames, Iowa 50011, United States}
\author{Ratul Chowdhury}
\email{ratul@iastate.edu}
\affiliation{Department of Chemical and Biological Engineering, Iowa State University, Ames, Iowa 50011, United States}

\begin{document}

\begin{tocentry}
\includegraphics{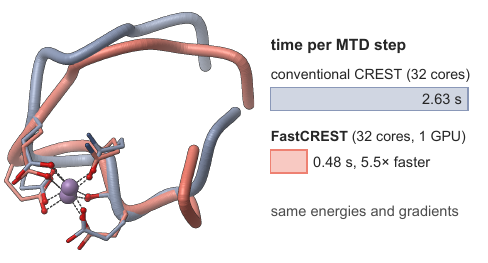}
\end{tocentry}

\begin{abstract}
CREST conformer searches with GFN2-xTB take days for large, flexible molecules: one search of a 274-atom gallium(III)--peptide complex with CREST 3.0.2 took 321.5~h on 32 CPU threads. FastCREST keeps CREST's algorithm and changes how a search uses the hardware. Its metadynamics (MTD) and molecular dynamics (MD) trajectories run with tuned OpenMP settings and one GPU, which solves the self-consistent-charge (SCC) eigenproblem and evaluates the three-body dispersion term. About 15,000 geometry optimizations per iteration run as Slurm array jobs on other CPU nodes, and checkpoints let a search continue on preemptible GPUs. On A100 nodes, the OpenMP settings and loop changes made an MTD step 2.4 times faster and the GPU another 2.3 times; array jobs made a batch of optimizations 39 times faster. For six more molecules of 24 to 339 atoms, a FastCREST step was 2.2 to 7.1 times faster than a conventional CREST step. A complete FastCREST search, which also lifts CREST~3's cap of 200 optimization cycles, took 54.3~h, evaluating energies and gradients 8.1 times as fast as the CREST 3.0.2 search (on different hardware) with fewer CPU-hours per evaluation. Two opt-in settings (a warm-started Jacobi eigensolver and GPU-resident SCC iterations), LLVM's OpenMP library and one CPU socket per trajectory made an MTD step 8.3 times faster than conventional CREST. Energies agree with CREST 3.0.2 within $1.01\times10^{-10}$~\Eh{} and gradients within $8.2\times10^{-11}$~\Eh~bohr$^{-1}$, and controlled MD runs passed a prespecified equivalence test. Most replicate FastCREST and conventional CREST searches ended in similar compact conformers, with no detectable difference in energy.
\end{abstract}

\section{Introduction}

Metal-binding peptides are candidates for the recovery of critical minerals, because their sequences can be engineered to carry metal-coordinating groups that raise the affinity for the desired metal and the selectivity over competing metal ions.\cite{braun2018,schoenberger2019,schoenberger2021} We design gallium-binding peptides to recover gallium selectively from industrial waste streams. A peptide can adopt many distinct conformations, and the arrangement of its backbone and side chains affects which coordinating groups can reach the metal, the geometry of the binding site and the relative stability of the metal--peptide complex. A geometry optimization from one starting structure finds only a local minimum and cannot capture this diversity. We therefore sample the conformers of each complex, compare how they coordinate gallium, and refine representative low-energy conformers with density functional theory.\cite{grimme2021}

CREST with the GFN2-xTB tight-binding method is a standard tool for generating conformer ensembles.\cite{crest2020,crest3,gfn2} Its iMTD-GC workflow (iterative metadynamics with genetic crossing)\cite{crest2020,grimme2019mtd} runs metadynamics (MTD)\cite{laio2002} trajectories biased by the root-mean-square deviation (RMSD) from earlier structures, optimizes thousands of their snapshots, and repeats until no lower conformer appears. One energy-and-gradient (E/G) evaluation takes a few seconds on one CPU core for a few hundred atoms, but a search needs millions of them. For a 274-atom gallium(III)--peptide complex in implicit water, one search with the CREST 3.0.2 release took 321.5~h (13.4 days) on 32 CPU threads, a cost that limits how many peptide candidates can be screened. More threads hardly help: even with the thread settings of Section~\ref{sec:threads}, 64 threads instead of 32 made an MTD time step only 1.06 to 1.09 times faster.

A CREST search mixes a few long trajectories, whose speed is set by the time of one E/G evaluation, with many thousands of independent geometry optimizations, and CREST runs both on the cores of one node. GPU implementations of semiempirical and tight-binding methods\cite{wu2012,maia2012,zhou2020,vuong2023,dxtb} speed up single calculations on large systems, but not this mix of short, sequential trajectory steps and many small independent jobs.

FastCREST changes how a search uses the hardware and leaves the energies and gradients unchanged (Figure~\ref{fig:method}). The trajectories run with OpenMP settings that let their threads share the work, and with one GPU for the eigensolver and the three-body dispersion term. The optimizations run as Slurm array jobs on other CPU nodes, and checkpoints let a search run on preemptible GPUs of a shared cluster. Apart from lifting a fixed cap on optimization cycles (Section~\ref{sec:search}), FastCREST's results differ from CREST's only through the order of floating-point operations and by less than the convergence threshold of the self-consistent-charge (SCC) procedure.

\section{Methods}
\label{sec:method}

\subsection{Conformer Search and Test System}
\label{sec:search}

\begin{figure*}[tbp]
\centering
\includegraphics[width=\textwidth]{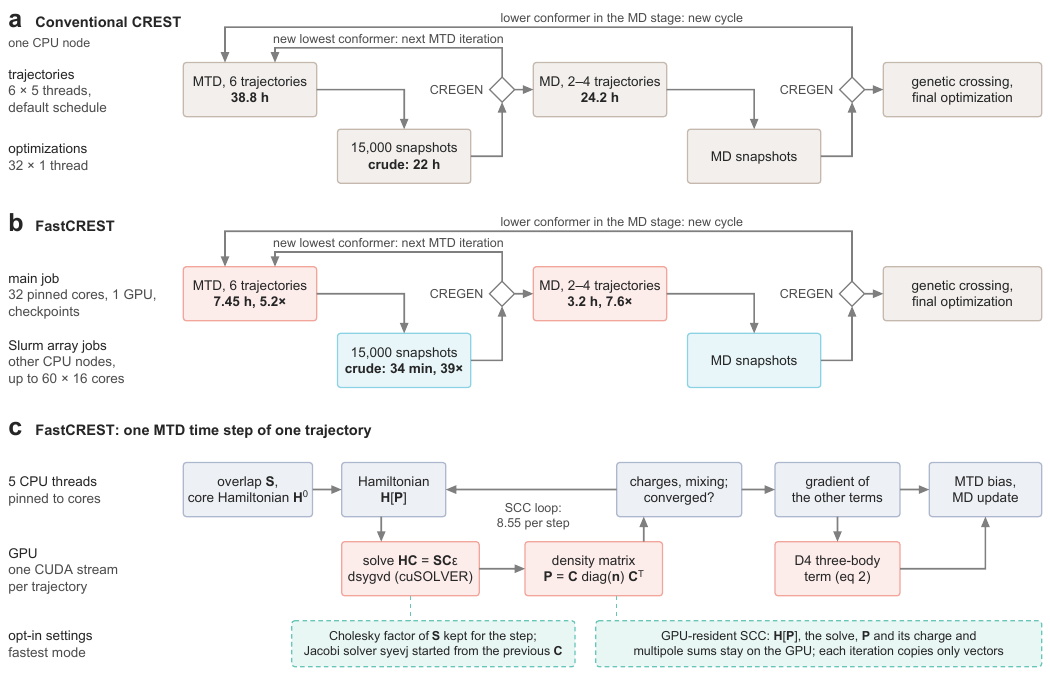}
\caption{How conventional CREST and FastCREST run the same iMTD-GC search. (a) Conventional CREST runs every stage on the threads of one CPU node; mean times per MTD iteration, per MD stage and per crude batch of optimizations in the CREST 3.0.2 search. (b) FastCREST runs the trajectories in a main job with one GPU (red) and checkpoints, and the optimizations as Slurm array jobs on other CPU nodes (cyan); the same times for the FastCREST search of Table~\ref{tab:runs} (crude batch from the search with the GPU eigensolver only), with how many times faster than in (a). (c) One time step of one MTD trajectory: work on its five CPU threads (blue) and on its CUDA stream (red); the SCC loop runs 8.55 times per step on average. Green, dashed: the opt-in settings (Sections~\ref{sec:jacobi} and \ref{sec:devscc}).}
\label{fig:method}
\end{figure*}

In CREST's iMTD-GC workflow (Figure~\ref{fig:method}a), each iteration runs six MTD trajectories at once and optimizes their roughly 15,000 snapshots, first with crude and then with regular thresholds, sorting the results with CREST's ensemble tool CREGEN. Iterations repeat while they find a new lowest conformer. An MD stage at 400 and 500~K then samples from the lowest conformers; if its optimized snapshots contain a lower conformer, a new cycle of MTD iterations starts. Genetic crossing (GC) and a final optimization end the search, and the ensemble keeps the conformers within 5~kcal/mol of the lowest (the window set by \quickflag).

The test system is a 15-residue peptide (\mbox{MLTREQIEELSKKLK}) in which Ga$^{3+}$ binds five oxygens: the backbone carbonyl oxygen and one carboxylate oxygen of Glu5, the side-chain amide oxygen of Gln6, and one carboxylate oxygen each of Glu8 and Glu9 (274 atoms, neutral N- and C-termini, charge $+4$). All iMTD-GC searches of the test system used GFN2-xTB,\cite{gfn2} ALPB implicit water\cite{alpb} and \quickflag, and started from the same helical structure, preoptimized with GFN2-xTB. For comparison we use one complete search with the CREST 3.0.2 release\cite{crest302} on 32 threads with default OpenMP settings, run earlier for another study on a node type that was not recorded; it spent 48.2\% of its 321.5~h in MTD, 36.7\% in optimizations and 15.0\% in MD. We call it the CREST 3.0.2 search. Conventional CREST means CREST without our GPU code: the CREST 3.0.2 release, CREST built without our patches, or our builds run on CPUs only (the conventional CREST replicate searches). Table~S1 lists the settings of all searches.

FastCREST is built from CREST commit cfdc301f,\cite{crestcfdc301f} 80 commits after the 3.0.2 release, with tblite 0.5.0\cite{tblite} and the patches described below. Only one of those commits can change results: CREST's geometry optimizer (ANCOPT) now stops an unconverged optimization exactly at its cycle limit, whereas the release finishes its current block of steps first (at a limit of 200 cycles, up to 239). At a limit of 200 cycles, this left 2 to 3 more of every 100 MD snapshots unconverged than the release, more than the 1 per 100 we had set as acceptable (Table~\ref{tab:tests}). We kept the commit, because structures that converged in both versions passed or failed CREST's topology check alike, and controlled MD against the release passed (Section~\ref{sec:same}). FastCREST also lifts a cap on the optimization cycles. ANCOPT sets the cycle limit from the size of the molecule, one, three or five cycles per atom for crude, regular and tight convergence thresholds and at least 200, as the optimizer of the xtb program did when it ran CREST's optimizations before version 3. In every CREST 3 version, including the current code, this limit is fixed at 200 cycles for all optimizations of a search (Section~S2). FastCREST keeps the size-dependent limit (274, 822 and 1,370 cycles for the test system), so optimizations that CREST stops unconverged after 200 cycles can still converge. The single-thread energies and gradients of our builds on CPUs are bitwise identical to those of CREST built without our patches. All replicate searches, FastCREST and conventional CREST alike, used the size-dependent cycle limit and the thread settings of Section~\ref{sec:threads}, and most used the distributed optimizations of Section~\ref{sec:dist}, which do not change the computation.

\subsection{Thread Settings and Loop Changes}
\label{sec:threads}

\begin{figure}[tbp]
\centering
\includegraphics[width=3.25in]{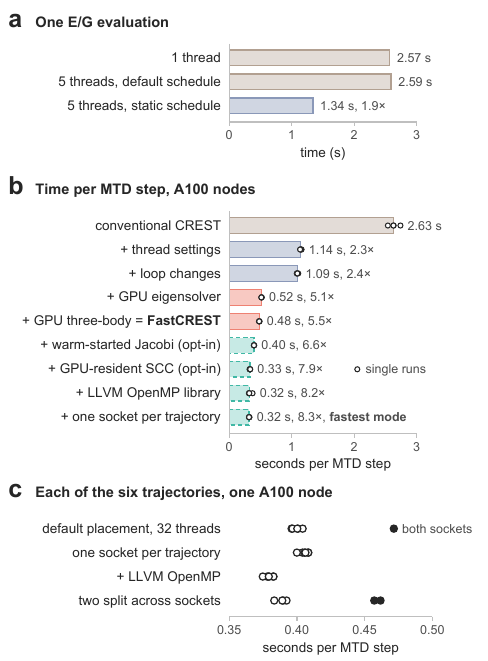}
\caption{Where the speed per MTD step comes from. (a) Time of one E/G evaluation of the test system on an AMD EPYC 7502 node with one thread, and with five threads under the default and a static OpenMP schedule, with the factor relative to the default schedule. (b) Time per MTD step as each change is added, with the factor relative to conventional CREST, on two A100 nodes (AMD EPYC 7543 hosts): median of four runs, two per node (bars), and the single runs (dots; one LLVM OpenMP run was slowed by other work on its node). All rows but the last used 32 of the job's 34 cores, the last 30 (one socket per trajectory). Dashed: opt-in settings. (c) Time per MTD step of each of the six trajectories (mean of two runs) with the Jacobi eigensolver, in a job with 16 and 24 cores on the two sockets: default thread placement, with one trajectory on both sockets (filled); one socket per trajectory, also with LLVM's OpenMP library; and a control with two trajectories on both sockets.}
\label{fig:speed}
\end{figure}

CREST runs the six trajectories of an MTD iteration at once, with five threads each in nested OpenMP parallel regions,\cite{openmp} and each optimization on one thread. The parallel loops of tblite take their schedule from the environment, and with default settings the GNU OpenMP library hands out one loop iteration at a time. Five threads then computed an E/G evaluation no faster than one (2.59 against 2.57~s), but in 1.34~s with a static schedule, 1.9 times faster (Figure~\ref{fig:speed}a). FastCREST's thread-settings script sets a static schedule, pins the threads of each trajectory to neighboring cores and enlarges the thread-stack cache of the C library, which made entering a nested parallel region 3.2 times faster (304 instead of 961~\textmu s). With the same binary, MTD steps then ran 2.2 to 2.3 times faster than with the settings of the CREST 3.0.2 search, on two A100 nodes and on CPU-only AMD EPYC 7502 nodes, where MD steps ran 2.0 to 2.5 times faster; optimizations did not change. We also replaced 43 uses of the Fortran intrinsic \texttt{spread} in tblite and its dependencies, which allocate a temporary array inside inner loops, with explicit loops in the same operand order; compiled without fused multiply--add, they round exactly as before.

Two further run-time settings change only the threading library and where the threads run. The GNU OpenMP library starts the threads of a nested team at every parallel region; LLVM's library (libomp 20.1.8), loaded in its place without recompiling, keeps them alive between regions. Single points agreed with those under the GNU library within the differences between repeated multithreaded runs ($1.7\times10^{-13}$~\Eh). When the cores of a job are split unevenly between the two sockets of a node, as in all our per-step measurements, the five cores of one trajectory can span both sockets; that trajectory then ran up to 18\% slower than the others (Figure~\ref{fig:speed}c), and every MTD iteration waits for it. Requesting 34 cores, which always include six groups of five cores that each lie on one socket, and binding each trajectory to one group (30 threads) removed the slow trajectory.

\subsection{GPU Eigensolver and Three-Body Dispersion}
\label{sec:gpu}

Each SCC iteration solves the generalized eigenproblem
\begin{equation}
\mathbf{H}[\mathbf{P}]\,\mathbf{C} = \mathbf{S}\,\mathbf{C}\,\boldsymbol{\varepsilon}, \qquad \mathbf{P} = \mathbf{C}\,\mathrm{diag}(\mathbf{n})\,\mathbf{C}^{\mathsf{T}},
\label{eq:gep}
\end{equation}
where the Hamiltonian $\mathbf{H}$ depends on the density matrix $\mathbf{P}$, $\mathbf{S}$ is the overlap matrix, $\mathbf{C}$ holds the orbital coefficients, $\boldsymbol{\varepsilon}$ the orbital energies and $\mathbf{n}$ the Fermi-smeared occupations. For the test system the matrices are $671\times671$, and a time step takes 8.55 SCC iterations on average. During MTD and MD, FastCREST solves eq~\ref{eq:gep} with the double-precision \texttt{dsygvd} routine of cuSOLVER\cite{cusolver} and forms $\mathbf{P}$ with cuBLAS,\cite{cublas} with one CUDA stream per trajectory (Figure~\ref{fig:method}c). A solve that does not converge on the GPU is repeated on the CPU with the LAPACK routine of conventional CREST. A deterministic double-precision GPU kernel evaluates the three-body (Axilrod--Teller--Muto) term of the D4 dispersion correction,\cite{axilrod1943,muto1943,caldeweyher2019}
\begin{equation}
\begin{gathered}
E^{(3)} = \sum_{A<B<C} C_9^{ABC}\, f_\mathrm{damp}\,
\frac{3\cos\theta_A\cos\theta_B\cos\theta_C + 1}{(r_{AB}\,r_{BC}\,r_{CA})^3}, \\
C_9^{ABC} = 5\,\bigl(C_6^{AB} C_6^{BC} C_6^{CA}\bigr)^{1/2},
\end{gathered}
\label{eq:atm}
\end{equation}
with interatomic distances $r$, the interior angles $\theta$ of each atom triangle, $C_6$ coefficients at zero charge, and the GFN2-xTB damping function $f_\mathrm{damp}$.\cite{gfn2} The kernel runs once per E/G evaluation over up to 3.4 million atom triples and reproduces the CPU gradient within $4.4\times10^{-18}$~\Eh~bohr$^{-1}$. Optimizations stay on the CPU: 32 optimizations sharing one GPU ran at only 0.87 to 0.89 times the rate on CPU cores, and CPU results can be checked bit for bit.

\subsection{Optional Warm-Started Jacobi Eigensolver}
\label{sec:jacobi}

Within one E/G evaluation, $\mathbf{S}$ is fixed and $\mathbf{H}$ changes little from one SCC iteration to the next. An opt-in setting therefore factorizes $\mathbf{S}$ (Cholesky) once per evaluation and solves the resulting standard eigenproblem with the Jacobi solver of cuSOLVER (\texttt{syevj}), started from the eigenvectors of the previous SCC iteration (Figure~\ref{fig:method}c); in the per-step measurements of Section~\ref{sec:step}, a solve took 2.8 Jacobi sweeps on average. On the 306 structures of the single-point test (Section~\ref{sec:tests}), energies agreed with those of the standard GPU solver within $3.4\times10^{-13}$~\Eh{} and gradients within $3.4\times10^{-14}$~\Eh~bohr$^{-1}$; in the per-step measurements, none of 133,158 solves had to fall back to the standard solver. The setting is off by default, because it was tested only on A100 GPUs and its gain depends on the size of the molecule (Table~\ref{tab:size}).

\subsection{Optional GPU-Resident SCC Iterations}
\label{sec:devscc}

In each SCC iteration of the GPU code path above, the CPU builds $\mathbf{H}$, the GPU solves eq~\ref{eq:gep}, and $\mathbf{C}$ and $\mathbf{P}$ return to the CPU, which computes from $\mathbf{P}$ the Mulliken charges, the atomic dipole and quadrupole moments and the band energy; for the test system about 18~MB cross between CPU and GPU per iteration. A second opt-in setting keeps the iteration on the GPU (Figure~\ref{fig:method}c). The core Hamiltonian, the overlap and the dipole and quadrupole integrals are copied to the GPU once per E/G evaluation. In each iteration the CPU then sends only the potential shifts; the GPU builds $\mathbf{H}$ and solves eq~\ref{eq:gep}; the CPU sets the Fermi occupations; and the GPU forms $\mathbf{P}$ and returns eleven sums per basis function, from which the CPU assembles the charges, moments and energy. Charge mixing and all interaction potentials stay on the CPU, and $\mathbf{C}$ and $\mathbf{P}$ are copied back once, after the last iteration, for the gradient. An E/G evaluation then moves 48 instead of 154~MB. The GPU kernels use the formulas of tblite with the same operation order within each matrix element; only the sums over basis functions run in a different order. On the 306 structures of the single-point test (Section~\ref{sec:tests}), energies agreed with those of the standard GPU code path within $3.4\times10^{-13}$~\Eh{} and gradients within $5\times10^{-15}$~\Eh~bohr$^{-1}$ (with both opt-in settings, within $2.8\times10^{-13}$~\Eh{} of the Jacobi solver without the GPU-resident SCC), as closely as the GPU and CPU code paths agree with each other. The setting works with both eigensolvers; open-shell calculations keep the standard GPU code path.

\subsection{Distributed Optimizations and Checkpoints}
\label{sec:dist}

The main job, which runs the trajectories on one GPU node, sends each batch of 128 or more optimizations to one Slurm job array\cite{slurm} in chunks of 64 structures, and the array jobs run the same binary on other CPU nodes (Figure~\ref{fig:method}b). Before merging, the main job reoptimizes up to 32 structures from array jobs with its own CPU model, which must agree bitwise, and up to 32 from array jobs with other CPU models, which must agree within $10^{-7}$~\Eh{} in energy and $10^{-4}$~\AA{} in coordinates; any mismatch reruns the batch or the affected chunks. Each trajectory writes an exact checkpoint every 40 steps and each optimized structure is recorded, so a search stopped during its MTD iterations resumes from its last checkpoint; an interrupted MD stage restarts from its beginning. With checkpoints on, each trajectory starts its first SCC procedure from a fresh guess; this is the only change in behavior that checkpoints cause. We did not test whether resumed searches are statistically equivalent to uninterrupted ones.

\subsection{Performance Measures}
\label{sec:measures}

Searches adapt their length, so we compare the E/G throughput $R = N/t_\mathrm{wall}$ and the allocated cost per evaluation,
\begin{equation}
K_\mathrm{CPU} = \frac{1}{N}\sum_j c_j t_j, \qquad K_\mathrm{GPU} = \frac{1}{N}\sum_j g_j t_j,
\label{eq:cost}
\end{equation}
where $N$ counts the E/G evaluations of the main job and all array jobs, $t_\mathrm{wall}$ is the elapsed time of the main job, and job $j$ was allocated $c_j$ hardware threads and $g_j$ GPUs for its elapsed time $t_j$. Times per MTD step come from one batch of six MTD trajectories of 400 steps for each configuration, run on one node in the order of Figure~\ref{fig:speed}b and then in reverse.

\subsection{Tests of the Computation}
\label{sec:tests}

We compared single points of FastCREST with those of the release and of CREST built without our patches on 306 structures, and the GPU code path with the CPU one. Trajectories cannot be compared bit for bit, because multithreaded sums are not reproducible:\cite{demmel2015} even two runs of the same binary were more than 0.5~\AA{} apart (all-atom RMSD) after a median of 6.5~ps. We therefore ran controlled MD, 20~ps from four starting structures at 400 and 500~K with identical initial velocities, and compared the fraction of frames whose bonding topology changed. Two configurations count as equivalent if the paired 90\% confidence interval of the mean difference over 24 or 44 trajectory pairs lies within $\pm0.15$ (two one-sided tests\cite{tost}). Whole searches are not reproducible either, so we compared groups of replicate searches at a point that every search reaches, the lowest crude-level energy after its first MTD iteration, with the Hodges--Lehmann shift and its exact confidence interval.\cite{hodges1963,bauer1972} All criteria were fixed before the runs, except the rule for the replicate comparison, which was fixed after some individual results had been seen but before any group statistic was computed. Sections~S2 and S3 give the details.

\section{Results and Discussion}

\subsection{Time per MTD Step}
\label{sec:step}

Figure~\ref{fig:speed}b adds the changes one at a time on two A100 nodes (AMD EPYC 7543 hosts). Conventional CREST took 2.63~s per MTD step there (2.8~s in the CREST 3.0.2 search). The thread settings and loop changes cut this to 1.09~s, a factor of 2.4, and the GPU eigensolver and three-body term to 0.48~s, another factor of 2.3, so a FastCREST step was 5.5 times faster on the same nodes. The Jacobi eigensolver reached 0.40~s (6.6 times faster than conventional CREST). The GPU-resident SCC iterations reached 0.42~s alone (6.3 times) and 0.33~s with the Jacobi solver (7.9 times), and LLVM's OpenMP library 0.32~s (8.2 times). With one socket per trajectory as well, the fastest mode (both opt-in settings, LLVM's OpenMP library and one socket per trajectory) made a step 8.3 times faster than with conventional CREST and 1.5 times faster than with FastCREST. With the GPU-resident SCC, the trajectory whose cores span both sockets was only about 5\% slower than the others, so one socket per trajectory made the mean step only 1.01 times faster. With the Jacobi solver alone, that trajectory was 18\% slower, and one socket per trajectory made the slowest trajectory, which sets the pace of an MTD iteration, 1.15 times faster (Figure~\ref{fig:speed}c).

Each step is still mostly CPU work (63.5\% in a profile of an earlier version), and six trajectories kept the GPU 83 to 87\% busy in FastCREST and 79 to 87\% in the fastest mode. A second GPU per search therefore gained only a factor of 1.2, and two searches sharing one GPU and the 32 cores raised the combined output by at most a factor of 1.12, while each ran 1.6 to 1.9 times slower.

We repeated the measurement of Figure~\ref{fig:speed}b for six more closed-shell molecules of 24 to 339 atoms, on one A100 node of the same type (Table~\ref{tab:size}). The thread settings made a step 2.1 to 3.0 times faster for every molecule. The GPU eigensolver gained more as the molecule grew, from a factor of 1.05 for caffeine to 2.8 for a 16-unit amylose chain, so a FastCREST step was 2.2 to 7.1 times faster than with conventional CREST. The Jacobi solver added a factor of 1.2 to 1.3 for the medium-sized molecules but only 1.02 and 1.09 for the smallest and the largest. LLVM's OpenMP library helped most for the smallest molecule, caffeine (a factor of 3.1). In the fastest mode a step was 6.9 to 11.7 times faster than with conventional CREST.

\subsection{Complete Searches}
\label{sec:perf}

Table~\ref{tab:runs} lists the complete searches. The FastCREST search found a lower conformer twice, so it ran three cycles instead of two; with the extra cycle and without the cap of 200 optimization cycles, it made 1.37 times as many E/G evaluations as the CREST 3.0.2 search. It took 54.3~h, so its throughput was 8.1 times that of the CREST 3.0.2 search. Per million evaluations it was allocated 1,231 CPU-hours plus 7.5 GPU-hours, against at least 1,939 CPU-hours, so it is cheaper whenever a GPU-hour costs less than about 94 hardware-thread-hours (47 core-hours). Optimizations fell from 36.7\% to 13.6\% of the run time, so the trajectories now take 86\% of it. Without a GPU, searches with the thread settings, two of them also with distributed optimizations, reached 3.3 to 4.7 times the throughput of the CREST 3.0.2 search, and with the warm-started Jacobi eigensolver three FastCREST searches reached 9.5, 10.0 and 9.5 times. Two searches in the fastest mode reached 12.4 and 12.7 times and were allocated 1,195 and 1,170 CPU-hours plus 4.9 and 4.8 GPU-hours per million evaluations.

\begin{table*}[tbp]
\centering
\caption{Complete Searches of the Test System}
\label{tab:runs}
\small
\setlength{\tabcolsep}{3.3pt}
\begin{tabular}{@{}llcrrrrr@{}}
\toprule
 & & & & & times & & \\
search & hardware & MTD/MD & wall (h) & $N$ ($10^6$) & faster & $K_\mathrm{CPU}$ & $K_\mathrm{GPU}$ \\
\midrule
CREST 3.0.2\textsuperscript{\emph{a}} & not recorded & 4/2 & 321.5 & 5.31 & 1.00$\times$ & 1,939 & 0 \\
CPU, one node\textsuperscript{\emph{b}} & Zen~4, 64 cores & 4/2 & 100.3 & 5.50 & 3.32$\times$ & 2,334 & 0 \\
CPU, distributed\textsuperscript{\emph{c}} & Zen~4 & 4/3 & 98.7 & 6.17 & 3.79$\times$ & 2,851 & 0 \\
CPU, GPU code off\textsuperscript{\emph{d}} & Zen~5 & 2/1 & 33.8 & 2.62 & 4.71$\times$ & 2,255 & 0 \\
GPU eigensolver only\textsuperscript{\emph{e}} & A100/EPYC 7502 & 4/2 & 52.0 & 5.38 & 6.28$\times$ & 1,574 & 9.7 \\
\textbf{FastCREST}\textsuperscript{\emph{f}} & A100/EPYC 7543 & 5/3 & 54.3 & 7.26 & 8.11$\times$ & 1,231 & 7.5 \\
FastCREST + Jacobi, search 1\textsuperscript{\emph{g}} & A100/EPYC 7543 & 4/3 & 36.9 & 5.77 & 9.48$\times$ & 1,194 & 6.4 \\
FastCREST + Jacobi, search 2\textsuperscript{\emph{g}} & A100/EPYC 7543 & 3/2 & 28.4 & 4.69 & 9.99$\times$ & 1,228 & 6.1 \\
FastCREST + Jacobi, search 3\textsuperscript{\emph{g}} & A100/EPYC 7543 & 4/2 & 36.8 & 5.74 & 9.46$\times$ & 1,315 & 6.4 \\
Fastest mode, search 1\textsuperscript{\emph{h}} & A100/EPYC 7543 & 5/3 & 35.2 & 7.17 & 12.36$\times$ & 1,195 & 4.9 \\
Fastest mode, search 2\textsuperscript{\emph{h}} & A100/EPYC 7543 & 6/4 & 42.7 & 8.97 & 12.73$\times$ & 1,170 & 4.8 \\
\bottomrule
\end{tabular}

\smallskip
\begin{minipage}{\textwidth}\raggedright\footnotesize
One search per row, from the same input and settings, except that the CREST 3.0.2 search capped optimizations at 200 cycles (Section~\ref{sec:search}): MTD iterations and MD stages, elapsed time of the main job, E/G evaluations $N$, how many times faster than the CREST 3.0.2 search in E/G evaluations per hour (the searches ran different numbers of iterations, so their wall times are not comparable), and allocated CPU- and GPU-hours per million evaluations ($K_\mathrm{CPU}$, $K_\mathrm{GPU}$; eq~\ref{eq:cost}). Hardware: GPU and host CPU of the main job (AMD CPUs). All but the first two searches ran their optimizations as Slurm array jobs on other nodes. \textsuperscript{\emph{a}}Release version with default OpenMP settings; CPU-hours are 32 threads $\times$ wall time, not a recorded allocation. \textsuperscript{\emph{b}}Thread settings and loop changes, no GPU. \textsuperscript{\emph{c}}Thread settings, distributed optimizations and checkpoints, without the loop changes. \textsuperscript{\emph{d}}FastCREST's binary with its GPU code off, counted among the conventional CREST replicates; one cycle, because CREGEN fell back to the helical input after its MTD iterations (Section~\ref{sec:struct}). \textsuperscript{\emph{e}}Earlier version: no loop changes, three-body term on the CPU. \textsuperscript{\emph{f}}The default changes of Sections~\ref{sec:threads}, \ref{sec:gpu} and \ref{sec:dist}. \textsuperscript{\emph{g}}With the warm-started Jacobi eigensolver (Section~\ref{sec:jacobi}). \textsuperscript{\emph{h}}Both opt-in settings (Sections~\ref{sec:jacobi} and \ref{sec:devscc}), LLVM's OpenMP library and one socket per trajectory (30 threads). EPYC 7502 is a Zen~2 CPU, EPYC 7543 a Zen~3 CPU.
\end{minipage}
\end{table*}

\subsection{Same Computation}
\label{sec:same}

Table~\ref{tab:tests} summarizes the tests. Single points on 306 structures agreed with the release within $1.01\times10^{-10}$~\Eh{} in energy and $8.2\times10^{-11}$~\Eh~bohr$^{-1}$ in gradient, two orders of magnitude inside our criterion, and the GPU and CPU code paths of FastCREST, including both opt-in settings, differed by at most $4.6\times10^{-13}$~\Eh. All 13 controlled MD comparisons, six of them with one or both opt-in settings, passed the equivalence test.

Replicate searches scatter widely in both codes. Of the 15 finished conventional CREST replicates, one ended with the lowest final conformer of all replicate searches, 12 ended 3.7 to 25.2~kcal/mol above it, and two ended more than 110~kcal/mol above it because of a fallback in CREST itself (Section~\ref{sec:struct}). After the first MTD iteration, no difference between FastCREST and conventional CREST was detected: the lowest energies of the 16 FastCREST searches were typically 4.1~kcal/mol higher than those of the 17 conventional CREST searches (Hodges--Lehmann shift; 95\% confidence interval $-4.9$ to $+11.8$~kcal/mol). A FastCREST disadvantage of up to 11.8~kcal/mol cannot be excluded, and showing equivalence within $\pm2$~kcal/mol would need more than a hundred searches per group. Final energies agree with this (medians: FastCREST $+11.6$, conventional CREST $+16.4$~kcal/mol above the lowest final conformer; post hoc shift $-2.7$~kcal/mol, 95\% confidence interval $-11.3$ to $+6.5$). Table~S5 lists the searches. The evidence that FastCREST computes the same energies and gradients and samples equivalent trajectories therefore rests on the code-level tests.

\begin{table*}[tbp]
\centering
\caption{Tests of the Computation}
\label{tab:tests}
\small
\begin{tabular}{@{}L{0.16\textwidth}L{0.24\textwidth}L{0.24\textwidth}L{0.28\textwidth}@{}}
\toprule
test & compared & criterion & result \\
\midrule
search setup & eight configurations & same parameters & same (flexibility measure 0.391, six MTD trajectories of 250~ps, 272 SHAKE constraints\cite{shake}) \\
single points & 306 structures, nine configurations & largest $|\Delta E| \le 10^{-8}$~\Eh, largest gradient difference $\le 10^{-6}$~\Eh~bohr$^{-1}$ & $1.01\times10^{-10}$~\Eh{} and $8.2\times10^{-11}$~\Eh~bohr$^{-1}$ from the release \\
optimizer stop at the cycle limit (CREST change after 3.0.2) & 200 MD snapshots, our builds against the release, both at 200 cycles & at most 1 more unconverged per 100 & 2 to 3 more per 100: limit exceeded; change kept (Section~\ref{sec:search}) \\
controlled MD & 13 comparisons of 24 or 44 trajectory pairs & 90\% confidence interval of the mean difference within $\pm0.15$ & all 13 equivalent, in MD snapshots and after optimization \\
replicate searches & lowest energy after the first MTD iteration, 16 FastCREST and 17 conventional CREST searches & difference: 95\% interval excludes 0; equivalence: 90\% interval within $\pm2$~kcal/mol & inconclusive: shift $+4.1$~kcal/mol, 95\% interval $-4.9$ to $+11.8$ \\
released code & clean clone, rebuilt & byte-identical output; test script passes & byte-identical; 17 of 17 tests passed \\
\bottomrule
\end{tabular}
\end{table*}

\begin{table*}[tbp]
\centering
\caption{Time per MTD Step for Molecules of Different Size}
\label{tab:size}
\small
\begin{tabular}{@{}lrrrrrr@{}}
\toprule
 & & & conventional CREST & \multicolumn{3}{c}{times faster than conventional CREST} \\
\cmidrule(l){5-7}
molecule & atoms & charge & (s per step) & FastCREST & + Jacobi & fastest mode \\
\midrule
caffeine & 24 & 0 & 0.165 & 2.2$\times$ & 2.2$\times$ & 11.7$\times$ \\
taxol & 113 & 0 & 0.581 & 3.1$\times$ & 4.1$\times$ & 7.0$\times$ \\
Ga(III)--peptide & 132 & $+3$ & 0.782 & 3.5$\times$ & 4.6$\times$ & 6.9$\times$ \\
Ga(III)--peptide & 174 & $-5$ & 1.575 & 5.2$\times$ & 6.8$\times$ & 9.5$\times$ \\
Ga(III)--peptide & 236 & $-8$ & 2.358 & 5.2$\times$ & 6.4$\times$ & 7.8$\times$ \\
test system\textsuperscript{\emph{a}} & 274 & $+4$ & 2.630 & 5.5$\times$ & 6.6$\times$ & 8.3$\times$ \\
amylose, 16 units & 339 & 0 & 3.948 & 7.1$\times$ & 7.7$\times$ & 9.0$\times$ \\
\bottomrule
\end{tabular}

\smallskip
\begin{minipage}{\textwidth}\raggedright\footnotesize
Time per MTD step with conventional CREST, and how many times faster FastCREST, FastCREST with the Jacobi eigensolver and the fastest mode were. One job per molecule on an A100 node with an AMD EPYC 7543 host, mean of two runs per setting, measured as in Figure~\ref{fig:speed}b. All molecules closed shell; GFN2-xTB, ALPB water, \quickflag. \textsuperscript{\emph{a}}Median of the four runs of Figure~\ref{fig:speed}b.
\end{minipage}
\end{table*}

\subsection{Final Structures and Energies}
\label{sec:struct}

CREST's topology check keeps only structures with the bonds of the input, so every final conformer keeps the five Ga--O bonds, and the searches differ mainly in the rest of the peptide. The lowest final conformer of the replicate searches came from a conventional CREST search. It lies 137.3~kcal/mol below the input and is compact (radius of gyration 6.33 instead of 7.56~\AA): the helix has unfolded, and the second carboxylate oxygen of Glu8 sits 2.33~\AA{} from Ga, a contact that the topology check does not count as a bond. These energies and the preference for compact folds are GFN2-xTB/ALPB results that we did not validate further.

\begin{figure*}[tbp]
\centering
\includegraphics[width=\textwidth]{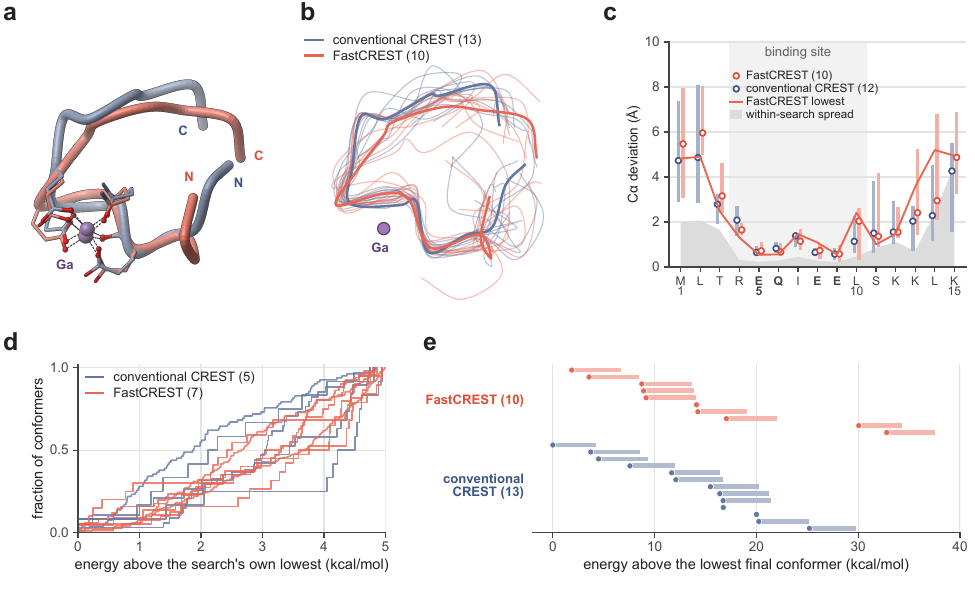}
\caption{Final conformers of the replicate FastCREST (red) and conventional CREST searches (blue); numbers in parentheses count searches. (a--c) Structures superposed on the binding site (Ga, C$\delta$ of Glu5, Gln6, Glu8 and Glu9, and the backbone of residues 4--10). (a) The lowest final conformer of the replicate searches (conventional CREST) and the lowest FastCREST conformer (best of ten searches, 1.9~kcal/mol higher); tubes trace the C$\alpha$ atoms, and dashes join Ga to oxygens within 2.6~\AA. (b) C$\alpha$ traces of the lowest conformer of each compact search, the two of (a) in bold. (c) Deviation of each residue's C$\alpha$ from the lowest final conformer, for the lowest conformer of each other compact search: medians (circles) and interquartile ranges (bars), the FastCREST conformer of (a) (line), and the largest deviation of any final conformer of the two searches in (a) from the lowest conformer of its own search (shaded). Bold residues bind Ga. (d) Cumulative fraction of conformers against the energy above the search's own lowest conformer, for ensembles with at least 10 conformers. (e) One row per compact search: its lowest conformer (dot) and the energy range of its final conformers (bar), relative to the lowest final conformer. Not shown in (b--e): two conventional CREST searches that ended near the helical input ($+116$ and $+125$~kcal/mol) and the FastCREST search with the Jacobi option ($+33.1$~kcal/mol).}
\label{fig:overlap}
\end{figure*}

Of the 26 finished replicate searches (10 FastCREST, 15 conventional CREST and one FastCREST search with the Jacobi option), all but two conventional CREST searches ended in compact conformers (heavy-atom radius of gyration at most 7.0~\AA, against 7.56~\AA{} for the input); we call these searches compact. In those two, the regular optimization of one MTD iteration changed the bonding topology of every structure in the energy window, and CREGEN fell back to the helical input (CREST code that is the same in the 3.0.2 release); both ran FastCREST's binary with its GPU code off. The lowest FastCREST conformer (best of ten searches) lies 1.9~kcal/mol above the lowest final conformer and has the same Glu8 contact (2.33~\AA). Superposed on the binding site\cite{kabsch1976} (Figure~\ref{fig:overlap}a), the two coincide within 2.4~\AA{} from Arg4 to Lys12, and both ends point elsewhere (Met1 to Thr3 and Lys13 to Lys15, 2.5 to 5.2~\AA). No other search ended with this conformer: the closest final conformer of any other search lies 1.2~\AA{} (heavy-atom RMSD) from it and 7.9~kcal/mol higher, and the lowest conformers of the other compact searches lie 0.8 to 3.7~\AA{} (backbone RMSD) from it (Figure~\ref{fig:overlap}b). FastCREST and conventional CREST searches deviate from it mainly at the two ends (Figure~\ref{fig:overlap}c): their interquartile ranges overlap at every residue, and their medians differ by more than 0.5~\AA{} at Met1, Leu2, Leu10, Leu14 and Lys15, where FastCREST's are 0.6 to 1.1~\AA{} larger.

Ensemble sizes vary widely (FastCREST 1 to 173 conformers, conventional CREST 1 to 82), because each search keeps only the conformers of its last cycle within 5~kcal/mol of its own lowest. Within that window, FastCREST and conventional CREST ensembles distribute their conformers alike: the median conformer lies 2.5 to 3.8~kcal/mol above the search's own lowest in FastCREST ensembles and 1.9 to 4.4~kcal/mol in conventional CREST ones (Figure~\ref{fig:overlap}d). Relative to the lowest final conformer, FastCREST searches scatter as widely as conventional CREST ones (lowest conformers 1.9 to 32.8 against 0 to 25.2~kcal/mol above it), and 2 of the 10 FastCREST and 2 of the other 12 compact conventional CREST searches ended within 5~kcal/mol of it (Figure~\ref{fig:overlap}e).

\subsection{Additional Applications and Benchmarks}

We also compared FastCREST with conventional CREST on other systems, with one CREST 3.0.2 release search per system (Sections~S7 to S9 and the last NCI-MTD row of Table~S7). For a second Ga(III)--peptide complex (264 atoms), FastCREST took 58.4~h on one A100 GPU with array jobs and conventional CREST 265.8~h on 32 threads of a CPU node, so FastCREST was 4.6 times faster, although it ran three cycles instead of two. Both final ensembles stayed close to the helical input, from which the last cycle of each search started again, and their lowest conformers lie within 0.7~\AA{} (C$\alpha$ RMSD) and 1.1~kcal/mol of each other with the same donor residues (Section~S8, Figure~S3(d-e)). FastCREST finished all four Zn--peptide systems (three 12-residue peptides, p1 to p3, and p1 also at low pH; Section~S7) in 13 to 34~h. Conventional CREST finished one, low-pH p1, in 97~h (FastCREST 7.3 times faster); we stopped its search of neutral p1 after 53~h, and its searches of p2 and p3 were not finished after 6~days (FastCREST at least 10 and 4.2 times faster). Where both codes gave structures, the peptide wraps around the Zn with mostly the same donors.

An NCI-MTD search of the test system (Ga(III)--peptide) took 34.2~h with FastCREST and 93.7~h with CREST 3.0.2 (FastCREST 2.7 times faster); with the wall potential of NCI-MTD, FastCREST runs the optimizations in the main job. The lowest-energy conformers obtained from both runs exhibited identical metal-coordinating residues and donor atoms.

A high-spin Fe(II)--peptide complex was also tested with fastCREST and CREST 3.0.2 (Sections~S9). The two searches exhibited differences in the lowest-energy conformers with distinct Fe(II) coordination environments, with CREST favoring a structure coordinated by four oxygen donors whereas FastCREST identified a more compact structure coordinated by six oxygen donors.
The FastCREST-derived minimum was 85.5 kcal mol \textsuperscript{-1} lower in GFN2-xTB energy than the final conventional CREST minimum (Figure S3 d,e; Table S9). Such differences may reflect the stochastic nature of conformational sampling \cite{king2024variant} and the greater electronic-structure complexity of open-shell transition-metal complexes, for which near-degenerate electronic states and self-consistent-field convergence difficulties have been reported in extended tight-binding calculations. \cite{moradi2026extensions}

Of CREST's other run types, protonation, deprotonation and tautomer searches gave the same lowest-energy protomers and tautomers with FastCREST as with conventional CREST; NCI-MTD and conformational entropy runs scatter from run to run in both codes, and their results agreed within that scatter where we repeated them (Table~S7).

\subsection{Limitations}
\label{sec:limits}

The replicate searches, the searches of Table~\ref{tab:runs} and most tests of the computation come from one system with \quickflag; the other metal--peptide systems were searched only once per code version (Section~\ref{sec:perf}). Per MTD step, the gains depended on the size of the molecule (Table~\ref{tab:size}) and will also depend on the CPU and the GPU (Table~S6). Most configurations in Table~\ref{tab:runs} ran one complete search, on different hardware, and the replicate groups ran under different conditions. Comparisons with the CREST 3.0.2 release or with CREST built without our patches (Table~\ref{tab:runs}, Table~S7 and Sections~S7 to S9) also include the 200-cycle cap that FastCREST lifts, which lets more of its optimizations converge and adds E/G evaluations; we did not measure its effect on the throughput or on the CPU-hours per evaluation. Of 40 FastCREST replicate searches started, 16 completed their first MTD iteration: a memory error on one GPU stopped 17 of them when they resumed, and, having seen interim results, we did not restart them; the other 7 were canceled early or still queued. 13 of these 16 FastCREST searches, but 7 of the 17 conventional CREST ones, resumed from checkpoints during their first iteration; 12 of the 16 finished it on L40S or V100 GPUs, which the code-level tests did not cover, and 6 computed the three-body term on the CPU for part of the run. Resuming in the first MTD iteration of a later cycle failed in four searches, in each case after the first MTD iteration of the search (fixed in version 1.2.0). In single searches of a high-spin Fe(II)--peptide complex with each code (264 atoms; Section~S9), most MTD and MD trajectories stopped early with these settings (48 of 50 in conventional CREST, all 20 in FastCREST), as in single FastCREST searches of Cu and Ni complexes of a related peptide (not shown); neither search sampled as intended, and the two ended in different structures. Although listed as a limitation of FastCREST, this behavior is identical in CREST. For ten structures of the Fe(II) complex, FastCREST's energies and gradients agreed with those of CREST 3.0.2 within $9\times10^{-11}$~\Eh{} and $1.2\times10^{-10}$~\Eh~bohr$^{-1}$. When CREST finds a lower conformer, it starts the next MTD trajectories from the optimized input structure, not from that conformer, and when this starts a new cycle it also discards the ensemble found so far (the 3.0.2 release does the same). This happened at least once in 24 of the 26 finished replicate searches, in both codes, so a final ensemble can lie above the lowest conformer seen during a search. FastCREST 1.3.0 can fix both (opt-in; Section~S6); Section~S7 gives an example with Zn-binding peptides.

\section{Conclusions}

On a GPU cluster, FastCREST ran CREST's iMTD-GC search of the test system with 8.1 times the E/G throughput of the CREST 3.0.2 search (12.4 and 12.7 times in the fastest mode), with unchanged energies and gradients. For the 274-atom test system, about half of the gain per MTD step comes from thread settings that any CREST user can apply and half from one GPU for the eigensolver and the three-body dispersion term; the GPU's share grows with the size of the molecule (Table~\ref{tab:size}). Running the optimizations as array jobs on other nodes cut their share of the run time from 37 to 14\%, so further speedup must come from the MTD step itself. Keeping the SCC iterations on the GPU made a step another 1.2 times faster with the Jacobi solver; the gradient, still computed on the CPU, is the next candidate. Caching the overlap factorization alone, sharing idle threads between trajectories, batching eigensolves in the optimizations, other compiler flags and Intel MKL gave little or no gain, and density-matrix purification would change the algorithm.

The importance of FastCREST lies in enabling more extensive conformational exploration of large and flexible molecular systems without changing the underlying GFN2-xTB level of theory used in conventional CREST. The close numerical agreement in energies and gradients, together with the identification of structurally similar and energetically competitive conformers, supports its use as an accelerated alternative for conformational screening. By substantially reducing computational cost, FastCREST enables investigation of larger systems, exploring multiple starting geometries, and faster screening of candidate molecules. Although individual searches may identify different energy minima due to the stochastic nature of conformational sampling, the increased computational efficiency enables broader exploration of conformational space and facilitates the selection of representative low-energy structures for subsequent higher-level calculations.

\section*{Data and Software Availability}

FastCREST is available at \url{https://github.com/chowdhurylab-studio/fastCREST} (version 1.3.0). It consists of patches to CREST cfdc301f,\cite{crestcfdc301f} tblite,\cite{tblite} dftd4,\cite{dftd4code} simple-dftd3,\cite{sdftd3code} multicharge\cite{multichargecode} and mctc-lib,\cite{mctclib} each under the license of the code it patches, with a build script that fetches the pinned upstream commits, the thread-settings script, Slurm job templates and a test script. With FastCREST's options switched off, single-thread energies and gradients are bitwise identical to those of CREST built without our patches. A clean build from GitHub reproduced the single-point output of the binary used here byte for byte and passed the test script (Table~\ref{tab:tests}). The input structures, the final ensembles and the raw outputs are available from the authors on request.

\begin{acknowledgement}
We thank Mahbuba Ferdaous for the structures of the open-shell metal complexes of the related peptide and Iowa State University Research IT for access to the Nova cluster. This research is partially supported by the U.S. Department of Energy, Office of Science, Office of Biological and Environmental Research, Biological Systems Science Division, 2025 Lab Critical Materials and Minerals (CMM) Pilot Program to R.C. Ames National Laboratory is operated for the U.S. Department of Energy by Iowa State University under Contract No. DEAC02-07CH11358. FastCREST model building was supported by the Critical Materials Innovation Hub, an Energy Innovation Hub funded by the US Department of Energy, Office of Critical Minerals and Energy Innovation (CMEI), Advanced Materials and Manufacturing Technologies Office (AMMTO) - under contract DEAC52–07NA27344 (LLNL-JRNL-857236). N.L.K was supported by Ames National Laboratory LDRD grant to design peptides for critical mineral metal binding, to R.C. The authors declare no competing financial interest.
\end{acknowledgement}

\begin{suppinfo}
Settings of all searches, software and hardware; details of the tests of the computation (single points, controlled MD, optimizer cycle limit); every replicate search; time per MTD step on other hardware; other CREST run types; CREST's new-cycle step and FastCREST's opt-in fixes; searches of Zn-binding test peptides; searches of Ga(III) and Fe(II) complexes of a second peptide (PDF).
\end{suppinfo}

\bibliography{refs}

\end{document}


\section{Search Settings, Software and Hardware}
\label{si:settings}

All iMTD-GC searches of the test system (274 atoms, charge $+4$, closed shell) started from the same structure and used
GFN2-xTB with ALPB water and \quickflag. FastCREST and conventional CREST searches were started with
\texttt{crest -{}-input search.toml -{}-newversion -{}-gfn2 -{}-chrg 4 -{}-uhf 0 -{}-alpb water -{}-quick -{}-ewin 5
-{}-noreftopo -T 32}; the input file sets the run type (\texttt{imtd-gc}), the calculation level and, for FastCREST,
\texttt{tblite\_solver = "cuda"}. The CREST 3.0.2 search used the equivalent command line of the release. In its command,
\quickflag{} followed \texttt{-{}-ewin 10} and reset the energy window to 5~kcal/mol, so the effective settings of all
searches are the same (Table~\ref{tab:si-settings}), except for the optimizer cycle limit (Section~\ref{si:cap}).

The CREST 3.0.2 release is the conda-forge build, which bundles a fork of tblite 0.3.0. Our builds (FastCREST, and CREST without our patches) used CREST commit cfdc301f with tblite 0.5.0, gfortran 11.5 (\texttt{-O3}, no \texttt{-march}, so no
fused multiply--add instructions), OpenBLAS 0.3.27 and CUDA 12.4.1. Of the 200 files under \texttt{src/}, 151 are
identical in the two CREST trees, including the MD and MTD code, CREGEN and every optimizer file except the one with the
cycle limit.

All jobs ran on the Nova cluster of Iowa State University under Slurm. GPU nodes held A100 (40 and 80~GB), H200, V100,
L40S or RTX PRO 6000 GPUs, with AMD EPYC 7413, 7502, 7513, 7543 or 9645, AMD Zen~5 or Intel Xeon hosts. CPU-only searches
ran on EPYC 7502, Zen~4 (EPYC 9654) and Zen~5 (EPYC 9645) nodes; the optimizations run as Slurm array jobs went to
whichever CPU nodes the scheduler assigned.

\begin{table}[tbp]
\caption{Effective Search Settings of the iMTD-GC Searches of the Test System}
\label{tab:si-settings}
\small
\begin{tabular}{@{}L{0.27\textwidth}L{0.68\textwidth}@{}}
\toprule
item & setting \\
\midrule
algorithm & iMTD-GC, \quickflag \\
MTD & 6 trajectories per iteration, 250~ps each at 300~K; RMSD bias with push strengths 0.548 and 0.274 and widths $\alpha$ = 1.2, 0.6 and 0.3~bohr$^{-2}$; bias structures stored every 1~ps \\
integration & leap-frog, 5~fs time step, hydrogen mass 2~u, Berendsen thermostat (500~fs), snapshots every 100~fs \\
bond constraints & SHAKE on bonds with GFN0 Wiberg bond order $>0.5$ and length within $1.2(r_i+r_j)$; metal--ligand bonds never constrained (272 constraints) \\
MD stage & lowest one or two conformers, at 400 and 500~K, 125~ps each \\
optimizations & MTD snapshots crude, then regular; MD snapshots regular; genetic crossing crude, then regular; final ensemble tight \\
convergence (energy, gradient) & crude $5\times10^{-4}$~\Eh, $1\times10^{-2}$~\Eh~bohr$^{-1}$; regular $5\times10^{-6}$, $1\times10^{-3}$; tight $1\times10^{-6}$, $8\times10^{-4}$ \\
optimizer & ANCOPT; cycle limit one, three or five cycles per atom for crude, regular and tight thresholds, at least 200 (the CREST 3.0.2 search: 200, which the release can exceed by up to 39); \texttt{hlow} 0.005; 20 micro-iterations \\
CREGEN & energy window 5~kcal/mol (twice that at the crude level); RMSD 0.125~\AA; energy 0.05~kcal/mol; rotational constants 1\%; topology check on \\
genetic crossing & at most 2,500 structures \\
SCC & energy $1\times10^{-6}$~\Eh, density $2\times10^{-5}$ \\
\bottomrule
\end{tabular}
\end{table}

\section{Tests of the Computation}
\label{si:tests}

\paragraph{Configurations.} The tests compare nine configurations: the CREST 3.0.2 release; conventional CREST with the
command-line input and with the input file; FastCREST with the CPU solver, alone and with the warm-started Jacobi
eigensolver switched on; and FastCREST on the GPU, alone, with the Jacobi eigensolver, with the GPU-resident SCC, and
with both opt-in settings. Every GPU calculation had to show GPU use (eigensolves and the three-body term on the GPU, without a CPU fallback); all did.

\paragraph{Search setup.} Eight configurations replayed the search up to its first MTD. All made the same decisions as the
CREST 3.0.2 search: the same preoptimized structure (energies of its first and last optimization steps $-415.2625315087$
and $-415.2625672259$~\Eh, the same to $1\times10^{-10}$~\Eh), flexibility measure 0.391, six MTD trajectories of
250~ps with the same bias settings, no failed trial MTD, 274 bonds with Wiberg bond order above 0.5 and 272 SHAKE
constraints.

\paragraph{Single points.} Energies and gradients were computed for 306 structures: the two input geometries, the 135
final conformers and 50 MD snapshots of the CREST 3.0.2 search, two steps of its first optimization, and 50 MD-stage
inputs and 67 conformers of a FastCREST search. The criterion was $|\Delta E| \le 10^{-8}$~\Eh{} and a largest gradient
component difference $\le 10^{-6}$~\Eh~bohr$^{-1}$; every structure met it in every comparison (Table~\ref{tab:si-sp}).
Our builds without the GPU gave bitwise identical results; the differences of up to $10^{-10}$~\Eh{} between them and the release come from
the different builds (tblite version, compiler and flags). The apo peptide and the Co, Cu and Ni complexes of a related peptide (241 or 242 atoms; the complexes open shell with one or two unpaired electrons) agreed as well: over ten structures each,
the release, FastCREST on the CPU and FastCREST on the GPU agreed within $1.1\times10^{-10}$~\Eh{} and
$7.4\times10^{-11}$~\Eh~bohr$^{-1}$.

\begin{table}[tbp]
\caption{Single Points of 306 Structures}
\label{tab:si-sp}
\small
\begin{tabular}{@{}L{0.46\textwidth}rrr@{}}
\toprule
 & bitwise & largest & largest $|\Delta g|$ \\
comparison & identical & $|\Delta E|$ (\Eh) & (\Eh~bohr$^{-1}$) \\
\midrule
conventional CREST vs the release & 0 & $1.01\times10^{-10}$ & $8.16\times10^{-11}$ \\
input file vs command line (conventional CREST) & 306 & 0 & 0 \\
FastCREST, CPU solver vs conventional CREST & 306 & 0 & 0 \\
FastCREST + Jacobi, CPU solver vs FastCREST, CPU solver & 306 & 0 & 0 \\
FastCREST, GPU vs the release & 0 & $1.01\times10^{-10}$ & $8.16\times10^{-11}$ \\
FastCREST + Jacobi, GPU vs the release & 0 & $1.00\times10^{-10}$ & $8.16\times10^{-11}$ \\
FastCREST + GPU-resident SCC vs the release & 0 & $1.00\times10^{-10}$ & $8.16\times10^{-11}$ \\
both opt-in settings vs the release & 0 & $1.00\times10^{-10}$ & $8.16\times10^{-11}$ \\
FastCREST, GPU vs CPU solver & 0 & $3.98\times10^{-13}$ & $4.0\times10^{-15}$ \\
FastCREST + Jacobi, GPU vs CPU solver & 0 & $3.41\times10^{-13}$ & $3.3\times10^{-14}$ \\
FastCREST + GPU-resident SCC vs CPU solver & 0 & $4.55\times10^{-13}$ & $4.0\times10^{-15}$ \\
both opt-in settings vs CPU solver & 0 & $4.55\times10^{-13}$ & $6.0\times10^{-15}$ \\
\bottomrule
\end{tabular}
\end{table}

\paragraph{Controlled MD.} Trajectories cannot be compared bit for bit, so we compared what matters for the search: how
often MD snapshots change their bonding topology, which decides how many structures CREGEN keeps. MD trajectories of
20~ps at 400 and 500~K started from four structures with the same initial velocities in every configuration (taken from
CREST restart files): three sets of initial velocities for two structures and eight for two structures of a later cycle, 44 trajectory pairs per comparison (24 when only the first three sets were run). Every tenth snapshot was optimized with the 200-cycle limit. For each pair
we took the difference $d$ in the fraction of snapshots whose topology differs from the input, before and after
optimization. A comparison passed if the 90\% confidence interval of the mean $d$ lay within $\pm0.15$ (two one-sided
tests; the margin was fixed before the runs). All 13 comparisons passed (Table~\ref{tab:si-md}). Paired trajectories
separated (all-atom RMSD above 0.5~\AA) after a median of 4.7 to 4.8~ps in every comparison with the release, and after
6.5~ps when the same build ran twice, so no configuration diverged early.

\begin{table}[tbp]
\caption{Controlled MD: 90\% Confidence Intervals of the Mean Difference in Topology-Change Fraction}
\label{tab:si-md}
\small
\begin{tabular}{@{}L{0.52\textwidth}rcc@{}}
\toprule
comparison & pairs & MD snapshots & after optimization \\
\midrule
conventional CREST run twice & 24 & $-0.024$ to $+0.017$ & $-0.084$ to $+0.041$ \\
conventional CREST vs the release & 44 & $-0.015$ to $+0.017$ & $-0.039$ to $+0.030$ \\
input file vs command line & 24 & $-0.005$ to $+0.034$ & $-0.030$ to $+0.110$ \\
FastCREST, CPU solver vs conventional CREST & 24 & $-0.043$ to $+0.005$ & $-0.137$ to $-0.007$ \\
FastCREST, GPU vs CPU solver & 24 & $-0.000$ to $+0.050$ & $+0.008$ to $+0.134$ \\
FastCREST, GPU vs conventional CREST & 44 & $-0.001$ to $+0.030$ & $+0.005$ to $+0.072$ \\
FastCREST, GPU vs the release & 44 & $-0.003$ to $+0.033$ & $-0.014$ to $+0.082$ \\
FastCREST + Jacobi vs FastCREST, GPU & 44 & $-0.033$ to $+0.010$ & $-0.076$ to $+0.025$ \\
FastCREST + Jacobi vs the release & 44 & $-0.020$ to $+0.026$ & $-0.048$ to $+0.064$ \\
FastCREST + GPU-resident SCC vs FastCREST, GPU & 44 & $-0.024$ to $+0.016$ & $-0.065$ to $+0.040$ \\
both opt-in settings vs FastCREST + Jacobi & 44 & $-0.021$ to $+0.021$ & $-0.059$ to $+0.040$ \\
FastCREST + GPU-resident SCC vs the release & 44 & $-0.005$ to $+0.027$ & $-0.029$ to $+0.072$ \\
both opt-in settings vs the release & 44 & $-0.013$ to $+0.020$ & $-0.044$ to $+0.042$ \\
\bottomrule
\end{tabular}
\end{table}

\paragraph{Optimizer cycle limit.}
\label{si:cap}
CREST commit 9908b42, between the 3.0.2 release and cfdc301f, stops an unconverged ANCOPT optimization exactly at its
cycle limit; the release lets it finish its last relaxation block, which at a limit of 200 cycles means up to 239. Unconverged optimizations are discarded, so the earlier stop can only remove structures. We optimized 100 MD snapshots of each of two sets at the regular level
(Table~\ref{tab:si-cap}). The earlier stop left 2 to 3 more of every 100 unconverged, more than the 1 per 100 we had set as acceptable; runs with a 240-cycle limit put an upper bound of 4 to 8 per 100 on the difference (Table~\ref{tab:si-cap}). Structures converged in both versions passed
or failed CREST's topology check alike, and their energies differed by at most $1.5\times10^{-3}$~\Eh, as optimizations
in different builds do. We kept the commit.

This test used a limit of 200 cycles in both versions, as CREST does in its searches. ANCOPT derives the limit from the
number of atoms when none is set (Table~\ref{tab:si-settings}), by the same rule as the optimizer of the xtb program,
which ran CREST's optimizations before version 3. In every CREST 3 version (3.0, 3.0.1, 3.0.2 and the current code),
however, the routine that writes the optimization settings to the output also sets the limit to 200 for the rest of the
search. FastCREST's version of that routine changes no setting, so each optimization keeps the size-dependent limit, and
structures that stay unconverged after 200 cycles can still converge. Every replicate search, FastCREST and
conventional CREST, used the size-dependent limit.

\begin{table}[tbp]
\caption{Optimizations Converged at a Limit of 200 Cycles, out of 100}
\label{tab:si-cap}
\small
\begin{tabular}{@{}L{0.42\textwidth}rrrr@{}}
\toprule
snapshots & release & our builds & difference & upper bound \\
\midrule
MD stage of the CREST 3.0.2 search, second cycle & 95 & 92 & 3 & 8 \\
MD stage of a FastCREST search, second cycle & 93 & 91 & 2 & 4 \\
\bottomrule
\end{tabular}
\end{table}

\section{Replicate Searches}
\label{si:replicates}

The comparison was fixed after some individual results had been seen but before any group statistic was computed. The compared quantity is the lowest energy after the crude
optimization of the first MTD iteration (14,994 snapshots, the same CPU optimizer in every search), taken from every
search that reached that point, finished or not, so that searches that needed fewer cycles, and therefore finished first, are not favored. The groups are FastCREST on
the GPU and conventional CREST, that is, our builds run on CPUs only, whose single points are bitwise identical to
those of CREST built without our patches. The statistic is the Hodges--Lehmann shift (FastCREST minus conventional CREST) with exact Mann--Whitney
confidence intervals. A difference is detected if the 95\% interval excludes zero; equivalence is shown if the 90\%
interval lies within $\pm2$~kcal/mol (under half the 5~kcal/mol energy window); otherwise the result is inconclusive.

Sixteen of 40 FastCREST searches and 17 of 18 conventional CREST searches have a value at that point (Table~\ref{tab:si-searches}); conventional 18 passed it, but its value was not kept. Most of the other FastCREST searches stopped within seconds on one GPU with a memory error
and were not restarted. The 16 that reached it ran their first iteration on 9 L40S, 3 V100, 3 A100 and 1 H200 GPU; 13 of
them, but 7 of the 17 conventional searches, resumed from checkpoints before that point. The shift was $+4.09$~kcal/mol
(95\% interval $-4.94$ to $+11.75$, achieved confidence level 0.951; 90\% interval $-3.58$ to $+10.13$; two-sided $p = 0.35$): no
difference detected and equivalence not shown. With only the four FastCREST searches on A100 or H200 GPUs the shift
was $+3.12$ ($-14.50$ to $+19.80$), and against only the nine conventional searches with FastCREST's own build on the
CPU, $+2.75$ ($-6.44$ to $+13.89$). The final energies of the finished searches (10 FastCREST, 15 conventional) gave a shift of $-2.70$~kcal/mol
(95\% interval $-11.27$ to $+6.54$), a comparison chosen after the data were seen.

\begin{table}[tbp]
\caption{Replicate Searches of the Test System. Energies in kcal/mol Relative to the Lowest Final Conformer of These Searches (Conventional 7, $-415.48139$~\Eh)}
\label{tab:si-searches}
\small
\input{si_searches.tex}

\smallskip
\begin{minipage}{\textwidth}\raggedright\footnotesize
First MTD iteration: lowest energy after its crude optimization (the quantity compared). Final: lowest energy of
the final ensemble. New cycles: cycles started after the MD stage or genetic crossing found a lower conformer. GPU: the
GPU of the first iteration; conventional CREST ran on CPUs. Conventional 18 has no first-iteration value: its first iteration was
replayed from a checkpoint and its crude optimization was not kept.
\end{minipage}
\end{table}

\section{Time per MTD Step on Other Hardware}
\label{si:hardware}

Table~\ref{tab:si-hardware} lists the time per MTD step of the test system measured in searches and benchmarks on other
nodes. The host CPU matters as much as the GPU, because most of each step is still CPU work.

\begin{table}[tbp]
\caption{Time per MTD Step of the Test System on Different Hardware}
\label{tab:si-hardware}
\small
\begin{tabular}{@{}L{0.50\textwidth}rrr@{}}
\toprule
 & & & times faster than \\
code and hardware & threads & s per step & the CREST 3.0.2 search \\
\midrule
CREST 3.0.2 search, node not recorded & 32 & 2.8 & 1.0$\times$ \\
conventional CREST with the thread settings, EPYC 7502 & 32 & 1.43--1.51 & 1.9$\times$ \\
same, whole EPYC 7502 node & 64 & 1.35--1.37 & 2.1$\times$ \\
FastCREST without GPU, EPYC 7502 & 64 & 1.26--1.27 & 2.2$\times$ \\
FastCREST without GPU, Zen~4 (shared node) & 64 & 0.93 & 3.0$\times$ \\
FastCREST without GPU, Zen~5 (shared node) & 64 & 0.74 & 3.8$\times$ \\
FastCREST, V100 and an older Intel Xeon\textsuperscript{\emph{a}} & 8 & 1.86 & 1.5$\times$ \\
FastCREST, RTX PRO 6000 and EPYC 9645\textsuperscript{\emph{a}} & 32 & 0.68 & 4.1$\times$ \\
FastCREST, A100 and EPYC 7543 & 32 & 0.46--0.49 & 5.7--6.0$\times$ \\
FastCREST, H200 and Zen~5 & 32 & 0.39 & 7.1$\times$ \\
\bottomrule
\end{tabular}

\smallskip
\begin{minipage}{\textwidth}\raggedright\footnotesize
Ranges: two runs or two nodes. The hosts differ between rows, so the factors combine the code and the hardware; on the
same A100 nodes FastCREST was 5.5 times faster than conventional CREST (main text, Figure~2b).
\textsuperscript{\emph{a}}No GPU code for this architecture in the build: the three-body term ran on the CPU. V100: one
partial trajectory.
\end{minipage}
\end{table}

\section{Other CREST Run Types}
\label{si:modes}

CREST's other run types use the same trajectory and optimization code, so FastCREST runs them as well. We compared each with CREST built from the same commit without our patches, which stops optimizations after 200 cycles (Section~\ref{si:cap}; Table~\ref{tab:si-modes}). Protonation, deprotonation and tautomer
searches have no MD and gave the same structures and energies. The NCI-MTD and entropy runs are adaptive and scatter
from run to run in both codes; where runs were repeated, their results agree within that scatter. The taxol entropy runs ran once per code. The NCI-MTD search of the test system was run once with FastCREST and once with the CREST 3.0.2 release instead (last NCI-MTD row). With the wall potential of NCI-MTD, FastCREST
runs the optimizations in the main job rather than as array jobs, so its gain is smaller than in iMTD-GC searches.

\begin{table}[tbp]
\caption{Other Run Types with FastCREST and Conventional CREST}
\label{tab:si-modes}
\small
\begin{tabular}{@{}L{0.17\textwidth}L{0.25\textwidth}L{0.52\textwidth}@{}}
\toprule
run type & molecule & result \\
\midrule
protonation, deprotonation, tautomers & uracil, caffeine, guanine (gas phase) & same lowest-energy protomers and tautomers in every run (fastest mode: largest $|\Delta E|$ $6.4\times10^{-6}$~\Eh) \\
protonation & Ga(III)--peptide (132 atoms, $+3$, ALPB water) & lowest protomer the same within $2\times10^{-6}$~\Eh{} in all runs; the number of structures kept varies from run to run in conventional CREST itself (53, 69 and 60) as in FastCREST (74 and 58) \\
NCI-MTD & water trimer (gas phase) & same lowest energy ($-15.235025$~\Eh) in all nine runs \\
NCI-MTD & Ga(III)--peptide (132 atoms, ALPB water, \quickflag) & lowest energies of six runs within 1.8~kcal/mol; conventional CREST 59.6 and 217.7~min, FastCREST 18.1 and 29.0~min (3.3 and 7.5 times faster in the two pairs of runs, different nodes), fastest mode 16.0 and 14.6~min (3.7 and 14.9 times faster) \\
NCI-MTD & test system (ALPB water) & one search each from the same structure: CREST 3.0.2 release 93.7~h (32 threads, EPYC 7502 node), FastCREST 34.2~h (A100 GPU, EPYC 7513 host; 2.7 times faster, with three MTD iterations against two); lowest energies 5.0~kcal/mol apart (FastCREST higher); 1 conformer with FastCREST, 7 with the release \\
entropy & $n$-octane (gas phase) & conformational entropy 8.12~cal~mol$^{-1}$~K$^{-1}$ (FastCREST), 8.13 (fastest mode), 8.20, 8.20 and 8.32 (conventional CREST); for this 26-atom molecule FastCREST took 14.8~min and conventional CREST 7.3~min (2.0 times slower) \\
entropy & taxol (113 atoms, ALPB water, \quickflag) & conformational entropy 9.65 (conventional CREST, 63.3~h), 9.89 (FastCREST, 46.5~h, 1.4 times faster) and 9.37~cal~mol$^{-1}$~K$^{-1}$ (fastest mode, 21.6~h, 2.9 times faster); different nodes \\
\bottomrule
\end{tabular}
\end{table}

\section{CREST's New-Cycle Step and FastCREST's Opt-In Fixes}
\label{si:cycle}

When CREST's iMTD-GC search finds a lower conformer, two things in its code (CREST 3.0.2 and cfdc301f; the search
routine that the command line runs) do not do what the code intends. First, the routine that registers the new lowest
conformer deletes every \texttt{crest\_rotamers} file before the caller saves the newest of them for the next merge, so
when the MD stage or genetic crossing starts a new cycle, the ensemble found so far is lost and the next cycle merges
only its own MTD iterations. Second, the new lowest conformer is written to the file \texttt{coord} as the start of the
next MTD trajectories, but this routine never reads it again: every MTD trajectory starts from the optimized input structure. CREGEN's
topology check, however, compares with \texttt{coord}.

Of the 26 finished replicate searches of the test system, 24 started at least one new cycle after the MD stage or genetic
crossing (41 new cycles in all), in conventional CREST and FastCREST alike (Table~\ref{tab:si-searches}); with the lower conformers found during MTD iterations, the next MTD trajectories restarted from the input structure 60 times. Because both codes share
this behavior, it does not affect the comparison between them, but it explains why a final ensemble can lie far above
the lowest conformer seen during a search.

{\sloppy FastCREST 1.3.0 fixes both steps on request. With \texttt{CREST\_KEEP\_ENSEMBLE=1} the newest ensemble is copied before
it is deleted and joins every later merge; with \texttt{CREST\_RESTART\_FROM\_BEST=1} the next MTD trajectories start from
the new lowest conformer. With both unset, FastCREST computes what it did before: in a search forced into a second cycle,
the conformers and the search log were identical to the previous version, and single points were byte-identical. With
both set, the forced search restarted from the new lowest conformer, kept the earlier ensemble, and resumed normally when killed at each of its 33 checkpoint steps.\par}

\section{Zn-Binding Test Peptides}
\label{si:metal}

Three 12-residue $\alpha$-helices with random sequences (no Cys or Pro, one His, at least one Asp or Glu) were built in
PyMOL: \mbox{DEGLAHSMAAFK} (p1), \mbox{GINGQALHAEYW} (p2) and \mbox{LKERTTEIHENI} (p3), with neutral termini, Asp and Glu deprotonated and
Lys and Arg protonated. A Zn$^{2+}$ ion was placed 2.05~\AA{} from the His NE2 (charge $+1$; 174 to 211 atoms). p1 was
also built with Asp and Glu protonated (low pH, charge $+3$). Each system was searched once with conventional CREST (the
3.0.2 release, 32 CPU cores), with FastCREST 1.2.0 and with FastCREST 1.3.0 with both opt-in fixes
(Section~\ref{si:cycle}); FastCREST ran as in the main text, on one A100 or H200 GPU. All searches used GFN2-xTB, ALPB
water, \quickflag{} and \texttt{-{}-noreftopo}; FastCREST used the size-dependent cycle limit
(Table~\ref{tab:si-settings}), conventional CREST the cap of 200 cycles. The Zn was left out of CREGEN's topology check
(\texttt{-{}-notopo} with its atom number), because the check counts metal--ligand contacts as bonds and removes every structure in which the Zn
changes partners: with the default check, a FastCREST search of p3 ended 119~kcal/mol above the lowest structure of
Table~\ref{tab:si-zn}, with the Zn still on the His.

\begin{figure}[tbp]
\centering
\includegraphics[width=\textwidth]{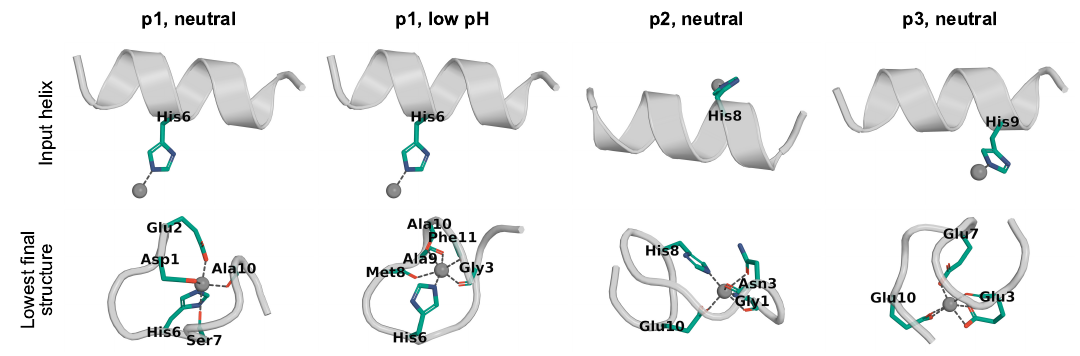}
\caption{Input helix (top) and lowest final structure (bottom; FastCREST with both fixes) of each peptide, all at the
same scale. Gray sphere: Zn; dashes: donors within 2.5~\AA; green sticks: the binding groups. Rendered with open-source
PyMOL.}
\label{fig:si-zn}
\end{figure}

In every other search that gave a structure, the peptide wraps around the Zn (Figure~\ref{fig:si-zn},
Table~\ref{tab:si-zn}). The helix unwinds, the Zn becomes six-coordinate (seven-coordinate in one structure), its first
and last donor residues lie 5 to 9 positions apart, and the radius of gyration falls from 6.6--6.8~\AA{} to
5.4--6.0~\AA. At neutral pH the donors are the His, Asp and Glu carboxylates and backbone carbonyls, in p2 also the
N-terminal amine and the Asn3 side chain; in p3 the carboxylates take five or six sites and the His is released. At low
pH the His binds with four or five backbone carbonyl oxygens.

FastCREST 1.2.0 finished every system in 13 to 34~h (42 to 79~h with both fixes, which add cycles). Conventional CREST
finished only low-pH p1, in 97~h (FastCREST 7.3 times faster); its searches of p2 and p3 were not finished after 6~days
(FastCREST at least 10 and 4.2 times faster). In neutral p1, the initial optimization of the input formed two bonds at the N-terminus that the input lacks. CREGEN compares with the input and removed all or nearly all structures of the MTD trajectories, which start from the optimized input, so valid structures came from the MD stages; with the restart fix, the MTD trajectories of
later cycles started from a valid structure. The conventional search of p1 had finished two MTD iterations without a valid
structure when we stopped it after 53~h. The finished conventional search ended 9.1~kcal/mol above the lowest structure
it had found, through the new-cycle step of Section~\ref{si:cycle}. With both fixes, FastCREST gave the lowest final
structure of any search for p1, p2 and p3. Each system was searched once per code; run-to-run scatter
(Section~\ref{si:replicates}) was not measured for these systems.

\begin{figure}[tbp]
\centering
\includegraphics[width=\textwidth]{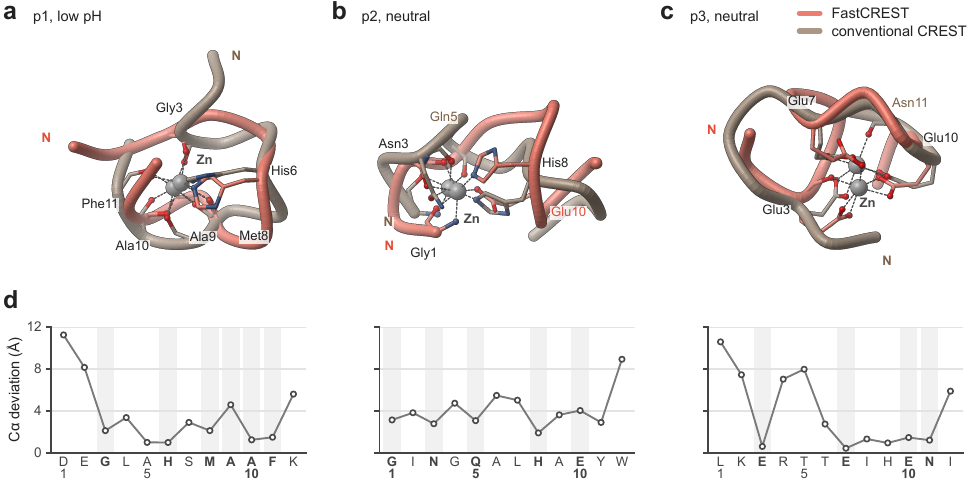}
\caption{Lowest structures found by FastCREST with both fixes (red) and by conventional CREST (brown) for (a) low-pH
p1, (b) p2 and (c) p3, superposed on the Zn site (the Zn, the backbone of the Zn-binding residues and their side-chain
donor groups). Dashes join the Zn (gray) to its donors; residue names in red or brown bind the Zn in only one
structure. (d) C$\alpha$ deviation per residue after this superposition; shaded residues bind the Zn in either
structure. Conventional CREST: for low-pH p1, the lowest structure its search found, which a new cycle dropped from its
final result; for p2 and p3, the lowest structure found before the 6-day limit. Neutral p1 is not shown, because
conventional CREST gave no structure.}
\label{fig:si-zn-overlap}
\end{figure}

Where both codes produced structures, they bind the Zn with mostly the same donors (Figure~\ref{fig:si-zn-overlap}).
Superposed on the Zn site, the lowest structures of the two codes lie 2.2 to 2.8~\AA{} apart (RMSD over the Zn, the
backbone of the binding residues and their donor groups) and share five or six donors; the two FastCREST searches of
each peptide differ from each other by 1.7 to 3.3~\AA{} on the same atoms. The chain ends differ more (C$\alpha$ RMSD
3.8 to 4.1~\AA).

\begin{table}[tbp]
\caption{Searches of the Zn-Binding Test Peptides. Energies in kcal/mol Relative to the Lowest Structure Found by Any
Search of That System}
\label{tab:si-zn}
\small
\setlength{\tabcolsep}{4pt}
\begin{tabular}{@{}L{0.10\textwidth}L{0.235\textwidth}rrL{0.37\textwidth}r@{}}
\toprule
system & search & time (h) & energy & Zn donors (CN) & $R_\mathrm{g}$ (\AA) \\
\midrule
p1 & FastCREST & 32.6 & $+2.5$ & His6; Glu2 (2 O); backbone O of Met8, Ala9, Ala10 (6) & 5.42 \\
 & FastCREST + both fixes & 41.8 & 0.0 & His6; Asp1 (2 O); Glu2; backbone O of Ser7, Ala10 (6) & 5.43 \\
 & conventional CREST\textsuperscript{\emph{a}} & 53 & & no valid structure & \\
p1, low pH & FastCREST & 13.3 & $+7.1$ & His6; backbone O of Gly3, Leu4, Ala9, Ala10, Phe11 (6) & 5.55 \\
 & FastCREST + both fixes & 43.6 & $+3.3$ & His6; backbone O of Gly3, Met8, Ala9, Ala10, Phe11 (6) & 5.50 \\
 & conventional CREST & 96.9 & $+9.1$ & His6; backbone O of Met8, Ala9, Ala10, Phe11; Met8 S (6) & 5.99 \\
 & \quad lowest found\textsuperscript{\emph{b}} & & 0.0 & His6; backbone O of Gly3, Met8, Ala9, Ala10, Phe11 (6) & 5.46 \\
p2 & FastCREST & 14.3 & $+2.8$ & His8; Glu10 (2 O); Asn3 OD1; backbone O of Gly1, Gln5 (6) & 5.72 \\
 & FastCREST + both fixes & 41.5 & 0.0 & His8; Glu10; Gly1 N and O; Asn3 O and OD1 (6) & 5.96 \\
 & conventional CREST\textsuperscript{\emph{c}} & $>$144 & $+1.6$ & His8; Gly1 N and O; Asn3 O and OD1; Gln5 O (6) & 5.79 \\
p3 & FastCREST & 34.4 & $+6.2$ & Glu3 (2 O); Glu7; Glu10 (2 O); Asn11 O (6) & 6.02 \\
 & FastCREST + both fixes & 79.1 & 0.0 & Glu3 (2 O and backbone O); Glu7; Glu10 (2 O) (6) & 5.62 \\
 & conventional CREST\textsuperscript{\emph{c}} & $>$144 & $+2.2$ & Glu3, Glu7, Glu10 (2 O each); Asn11 O (7) & 5.73 \\
\bottomrule
\end{tabular}

\smallskip
\begin{minipage}{\textwidth}\raggedright\footnotesize
FastCREST: version 1.2.0; with both fixes: version 1.3.0. Time: elapsed time of the main job, summed over the jobs the
search ran in. Energy, donors (within 2.5~\AA{} of Zn; CN, coordination number) and radius of gyration $R_\mathrm{g}$ of
the peptide heavy atoms: lowest final structure; $R_\mathrm{g}$ of the inputs: 6.68 (p1), 6.62 (p2) and 6.80~\AA{} (p3).
\textsuperscript{\emph{a}}Stopped by us after two MTD iterations. \textsuperscript{\emph{b}}Lowest structure the
conventional search found; a new cycle dropped it from the final result. \textsuperscript{\emph{c}}Not finished after
6~days: lowest structure found.
\end{minipage}
\end{table}

\section{A Second Ga(III)--Peptide Complex}
\label{si:ga2}

The peptide \mbox{MLTREQIEELSKLLE} differs from the test system in two residues (Leu13 and Glu15 instead of Lys). Its Ga(III)
complex has 264 atoms and charge $+1$ and is closed shell, with neutral termini, the four Glu deprotonated and Arg4 and
Lys12 protonated (neutral pH). In the preoptimized helical input, Ga binds six oxygens (Table~\ref{tab:si-pep023}). The complex
was searched once with each code from this input, with GFN2-xTB, ALPB water, \quickflag{} and \texttt{-{}-noreftopo} as
for the test system (final energy window 5~kcal/mol): conventional CREST (CREST 3.0.2) on 32 threads of an AMD Zen~5
node, and FastCREST 1.1.1 on one A100 GPU with an AMD EPYC 7543 host, with its optimizations run as 1,583 Slurm array
tasks on other nodes (6,060 allocated CPU-hours).
The two codes also used different optimizer cycle limits (Section~\ref{si:cap}): 200 cycles in CREST 3.0.2, and 264, 792
and 1,320 cycles for crude, regular and tight thresholds in FastCREST. In the two MD stages, 3,013 of 5,061 and 3,220 of
5,028 optimizations converged in CREST 3.0.2, and all MD-stage optimizations in FastCREST.
FastCREST took 58.4~h and conventional CREST 265.8~h, so FastCREST was 4.6 times faster, on different hardware and with
more work: it ran three cycles (six MTD iterations) and conventional CREST two (four). Per MTD iteration, FastCREST took
7.2~h and conventional CREST 35.2~h (4.9 times faster). No trajectory stopped early in either search.

Both searches ended near the input (Figure~\ref{fig:si-pep023}a--c). As in the test system, every structure kept the
Ga--O bonds of the input, and every conformer of both final ensembles kept the nine helical residues of the input and
lies within 1.5~\AA{} of it (C$\alpha$ RMSD). The two lowest final conformers coincide on the Ga site (RMSD 0.08~\AA) and
differ by 0.7~\AA{} over all C$\alpha$ atoms, mostly at Met1; FastCREST's lies 1.1~kcal/mol lower. The final ensembles
overlap: each of the 79 conventional CREST conformers lies within 1.1~\AA{} (C$\alpha$ RMSD) of one of the 48 FastCREST
conformers. Both searches had found lower structures that a new cycle then dropped (Section~\ref{si:cycle}). Conventional
CREST's lies 2.4~kcal/mol below FastCREST's lowest final conformer and only 0.2~\AA{} from it (C$\alpha$ RMSD).
FastCREST's, found in its second cycle, lies 30.7~kcal/mol below it with the same Ga--O bonds, but the helix has
unfolded into a compact structure (radius of gyration 6.7~\AA, input 7.5~\AA).
FastCREST's final ensemble comes from its third cycle, whose trajectories started again from the input. With one search
per code, run-to-run scatter (Section~\ref{si:replicates}) was not measured for this complex.

\section{An Open-Shell Fe(II) Complex}
\label{si:fe}

The high-spin Fe(II) complex of the same peptide (264 atoms, charge 0, four unpaired electrons, with the protonation
states and neutral termini of Section~\ref{si:ga2}) was searched once with each code, with the codes and settings of
Section~\ref{si:ga2}, from one preoptimized helical input in which Fe binds three oxygens. Conventional CREST took 242.3~h on 32 threads of an
Intel Skylake node, and FastCREST 7.3~h on one A100 GPU with an AMD EPYC 7513  host (63 array tasks, 222 CPU-hours). Both
searches finished normally, but most trajectories stopped early in both codes: 48 of 50 MTD and MD trajectories in
conventional CREST and all 20 in FastCREST, and the MTD iterations produced only 2 to 29\% of the snapshots of a complete
iteration. Neither search therefore sampled as intended. The two times are not comparable: conventional CREST ran four
cycles and FastCREST one, both with truncated trajectories.

\begin{figure}[tbp]
\centering
\includegraphics[width=\textwidth]{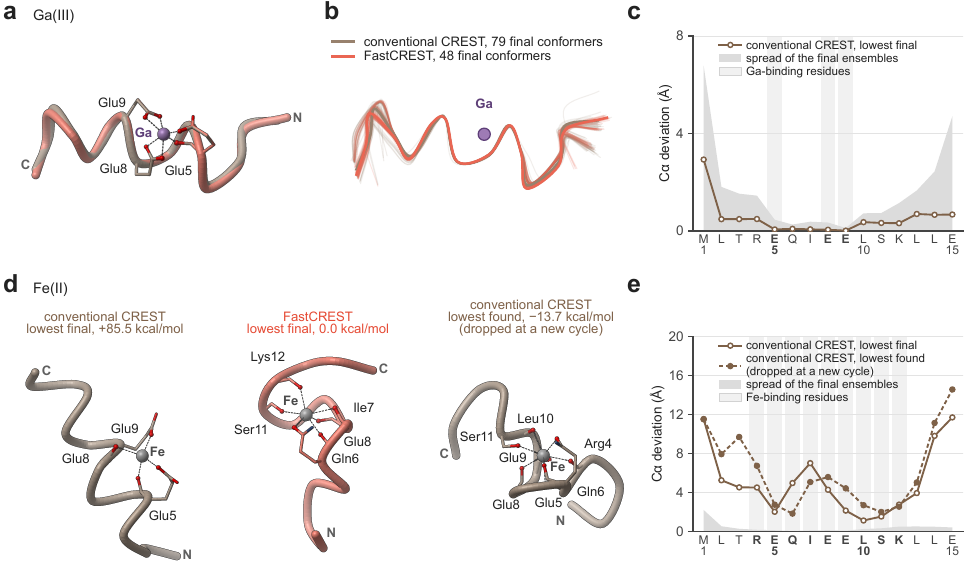}
\caption{Lowest structures of the Ga(III) (a--c) and high-spin Fe(II) complexes (d, e) of \mbox{MLTREQIEELSKLLE}, one search
per code; FastCREST red, conventional CREST brown. Every structure is superposed on the metal site of FastCREST's lowest
final conformer: the metal, the backbone of each residue that binds it in any structure compared (Ga: Glu5, Glu8, Glu9;
Fe: Arg4 to Lys12) and the binding side-chain atoms, a carboxylate as its C$\delta$. Tubes trace the C$\alpha$ atoms,
sticks show the binding groups, and dashes join the metal to oxygens within 2.6~\AA{} (Ga) or 2.8~\AA{} (Fe). (a) Lowest
final conformers (FastCREST's 1.1~kcal/mol lower). (b) C$\alpha$ traces of all final conformers, the lowest in bold. (c) Deviation of each
residue's C$\alpha$ from FastCREST's lowest final conformer, for conventional CREST's (solid) and for the dashed structure
of (b), and the largest deviation of any conformer in (b) from the lowest conformer of its own search (shaded); bold
residues belong to the site. The lowest structure conventional CREST found deviates by at most 0.8~\AA{} and is not
shown. (d) Side by side, in the same orientation and scale: the lowest final conformers of conventional CREST and
FastCREST (85.5~kcal/mol lower), and the lowest structure conventional CREST found (13.7~kcal/mol below FastCREST's
lowest final conformer, dropped at a new cycle). (e) As (c), for the two conventional CREST structures of (d).}
\label{fig:si-pep023}
\end{figure}

The two searches ended in different structures (Figure~\ref{fig:si-pep023}d,e; Table~\ref{tab:si-pep023}). Conventional
CREST's lowest final conformer stayed close to the input (C$\alpha$ RMSD 1.3~\AA) and binds Fe through the three oxygens
of the input and a carboxylate oxygen of Glu5. FastCREST's is compact and binds Fe through five backbone carbonyl oxygens
and the side-chain oxygen of Gln6, with no carboxylate; it lies 85.5~kcal/mol lower. In its third cycle, conventional
CREST had found a structure 13.7~kcal/mol below FastCREST's lowest final conformer, also compact and without a
carboxylate donor, but a new cycle dropped it (Section~\ref{si:cycle}), and the final ensemble ended 99.2~kcal/mol above
it. Single points gave the same energies with CREST 3.0.2 and FastCREST for the final conformers of both searches
(within $9\times10^{-11}$~\Eh), so the energy differences come from the different structures. The
structures and energies are GFN2-xTB results that we did not validate further.

\begin{table}[tbp]
\caption{Searches of the Ga(III) and High-Spin Fe(II) Complexes of \mbox{MLTREQIEELSKLLE}. Energies in kcal/mol Relative to
FastCREST's Lowest Final Conformer of Each Complex}
\label{tab:si-pep023}
\small
\setlength{\tabcolsep}{3pt}
\begin{tabular}{@{}L{0.08\textwidth}L{0.215\textwidth}rrrrL{0.245\textwidth}r@{}}
\toprule
complex & structure & time (h) & cycles & conformers & energy & metal donors (CN) & $R_\mathrm{g}$ (\AA) \\
\midrule
Ga(III) & input & & & & & Glu5; Glu8, Glu9 (2 O each); backbone O of Glu5 (6) & 7.49 \\
 & FastCREST & 58.4 & 3 & 48 & 0.0 & same as input (6) & 7.66 \\
 & \quad lowest found\textsuperscript{\emph{a}} & & & & $-30.7$ & same as input (6) & 6.71 \\
 & conventional CREST & 265.8 & 2 & 79 & $+1.1$ & same as input (6) & 7.36 \\
 & \quad lowest found\textsuperscript{\emph{a}} & & & & $-2.4$ & same as input (6) & 7.65 \\
Fe(II) & input & & & & & Glu9; backbone O of Glu5, Glu8 (3) & 7.57 \\
 & FastCREST\textsuperscript{\emph{b}} & 7.3 & 1 & 5 & 0.0 & Gln6 OE1; backbone O of Gln6, Ile7, Glu8, Ser11, Lys12 (6) & 6.57 \\
 & conventional CREST & 242.3 & 4 & 4 & $+85.5$ & Glu5, Glu9; backbone O of Glu5, Glu8 (4) & 7.25 \\
 & \quad lowest found\textsuperscript{\emph{a}} & & & & $-13.7$ & Gln6 OE1, Ser11 OG; backbone O of Arg4, Glu5, Gln6, Glu8, Glu9, Leu10 (8) & 6.45 \\
\bottomrule
\end{tabular}

\smallskip
\begin{minipage}{\textwidth}\raggedright\footnotesize
Time: elapsed time of the main job. Conformers: final ensemble. Energy, donors (within 2.6~\AA{} of Ga or 2.8~\AA{} of
Fe; CN, coordination number) and radius of gyration $R_\mathrm{g}$ of the peptide heavy atoms: lowest final conformer,
or the structure named; a Glu without a count binds one carboxylate oxygen. \textsuperscript{\emph{a}}Lowest structure
the search found; a new cycle dropped it from the final result (Section~\ref{si:cycle}); energy from the search log.
\textsuperscript{\emph{b}}Its lowest final conformer is also the lowest structure it found.
\end{minipage}
\end{table}

%% file: si_searches.tex
\begin{tabular}{@{}llrrrr@{}}
\toprule
search & GPU & first MTD iteration & final & conformers & new cycles \\
\midrule
conventional 1 & CPU & $+$13.1 & $+$16.7 & 19 & 1 \\
conventional 2 & CPU & $+$16.8 & $+$16.4 & 8 & 1 \\
conventional 3 & CPU & $+$21.1 & $+$25.2 & 3 & 2 \\
conventional 4 & CPU & $+$22.3 & not finished &  &  \\
conventional 5 & CPU & $+$24.5 & $+$4.5 & 7 & 3 \\
conventional 6 & CPU & $+$26.4 & not finished &  &  \\
conventional 7 & CPU & $+$29.4 & $+$0.0 & 3 & 1 \\
conventional 8 & CPU & $+$29.5 & $+$7.6 & 2 & 2 \\
conventional 9 & CPU & $+$32.1 & $+$20.2 & 12 & 1 \\
conventional 10 & CPU & $+$32.7 & $+$16.7 & 1 & 1 \\
conventional 11 & CPU & $+$34.0 & $+$15.5 & 12 & 2 \\
conventional 12 & CPU & $+$34.7 & $+$11.7 & 34 & 2 \\
conventional 13 & CPU & $+$35.0 & not finished &  &  \\
conventional 14 & CPU & $+$38.0 & $+$124.7 & 11 & 0 \\
conventional 15 & CPU & $+$41.1 & $+$116.0 & 10 & 0 \\
conventional 16 & CPU & $+$53.3 & $+$3.7 & 82 & 2 \\
conventional 17 & CPU & $+$54.7 & $+$20.0 & 1 & 3 \\
conventional 18 & CPU & -- & $+$12.1 & 9 & 1 \\
FastCREST 1 & H200 & $+$17.2 & not finished &  &  \\
FastCREST 2 & L40S & $+$18.2 & $+$3.6 & 22 & 2 \\
FastCREST 3 & L40S & $+$22.5 & not finished &  &  \\
FastCREST 4 & L40S & $+$23.2 & $+$1.9 & 51 & 1 \\
FastCREST 5 & A100 & $+$26.6 & $+$8.7 & 3 & 2 \\
FastCREST 6 & L40S & $+$31.4 & $+$32.8 & 40 & 2 \\
FastCREST 7 & L40S & $+$34.0 & not finished &  &  \\
FastCREST 8 & L40S & $+$34.7 & $+$14.1 & 1 & 2 \\
FastCREST 9 & L40S & $+$36.2 & $+$30.0 & 5 & 1 \\
FastCREST 10 & V100 & $+$37.1 & not finished &  &  \\
FastCREST 11 & V100 & $+$40.9 & not finished &  &  \\
FastCREST 12 & A100 & $+$42.1 & $+$9.2 & 20 & 3 \\
FastCREST 13 & V100 & $+$42.2 & not finished &  &  \\
FastCREST 14 & L40S & $+$47.7 & $+$8.9 & 173 & 1 \\
FastCREST 15 & A100 & $+$49.5 & $+$14.2 & 103 & 2 \\
FastCREST 16 & L40S & $+$50.8 & $+$17.1 & 19 & 1 \\
FastCREST + Jacobi & A100 & $+$37.9 & $+$33.1 & 10 & 2 \\
\bottomrule
\end{tabular}